\newif\iftwocolumns
\twocolumnsfalse     

\iftwocolumns
  \documentclass[aps,prl,twocolumn,groupedaddress,amsmath,amssymb,showpacs,floatfix]{revtex4}
\else
  \documentclass[aps,prl,preprint,groupedaddress,amsmath,amssymb,showpacs,floatfix]{revtex4}
\fi

\usepackage{graphicx}
\usepackage{dcolumn}
\usepackage{bm}
\usepackage{color}

\def\const{\mathop {\rm const}\nolimits } 
\def\F{\mbox{\rm F}} 
\def\E{\mbox{\rm E}} 
\def\K{\mbox{\rm K}} 

\def\const{\mathop {\rm const}\nolimits } 

\begin{document}



\title{Forceless Sadowsky strips are spherical}

\author{E.~L.~Starostin}
\email{e.starostin@ucl.ac.uk}
\author{G.~H.~M.~van~der~Heijden}
\email{g.heijden@ucl.ac.uk}
\affiliation{Department of Civil, Environmental \& Geomatic Engineering, \\
University College London, Gower Street, London WC1E 6BT, UK}

\date{\today}

\begin{abstract}
We show that thin rectangular ribbons, defined as energy-minimising
configurations of the Sadowsky functional for narrow developable elastic
strips, have a propensity to form spherical shapes in the sense that
forceless solutions lie on a sphere. This has implications for ribbonlike
objects in (bio)polymer physics and nanoscience that cannot be described
by the classical wormlike chain model. A wider class of functionals with
this property is identified.
\end{abstract}


\maketitle


\section{Introduction}
\label{intro}

Understanding the configurations and stresses of biopolymers lying on a
surface is important in a number of biomolecular processes,
including the packing of DNA inside viral capsids \cite{Jiang06}, cytokinesis
in animal and yeast cells during which mainly membrane-bound actin filaments
provide the forces necessary for cell division \cite{Kamasaki07},
and cell wall synthesis in bacteria \cite{Allard09,Salje11}.
Graphene nanoribbons have also been studied on surfaces \cite{Yin13}
with a view to assembling ribbon-like nanomaterials with desirable properties.

A classical theoretical approach to the study of such filamentous objects
is to use the wormlike chain (WLC) model
\cite{Marko95} 
in which the polymer is assumed to have only entropic bending
elasticity (characterising the persistence length). For biopolymers, like
DNA, that also have torsional elasticity, the torsional directed walk or
rodlike chain (RLC) is a more appropriate model \cite{Marko94a,Moroz97}.

\begin{figure}
\begin{center}
\includegraphics[width=0.6\linewidth]{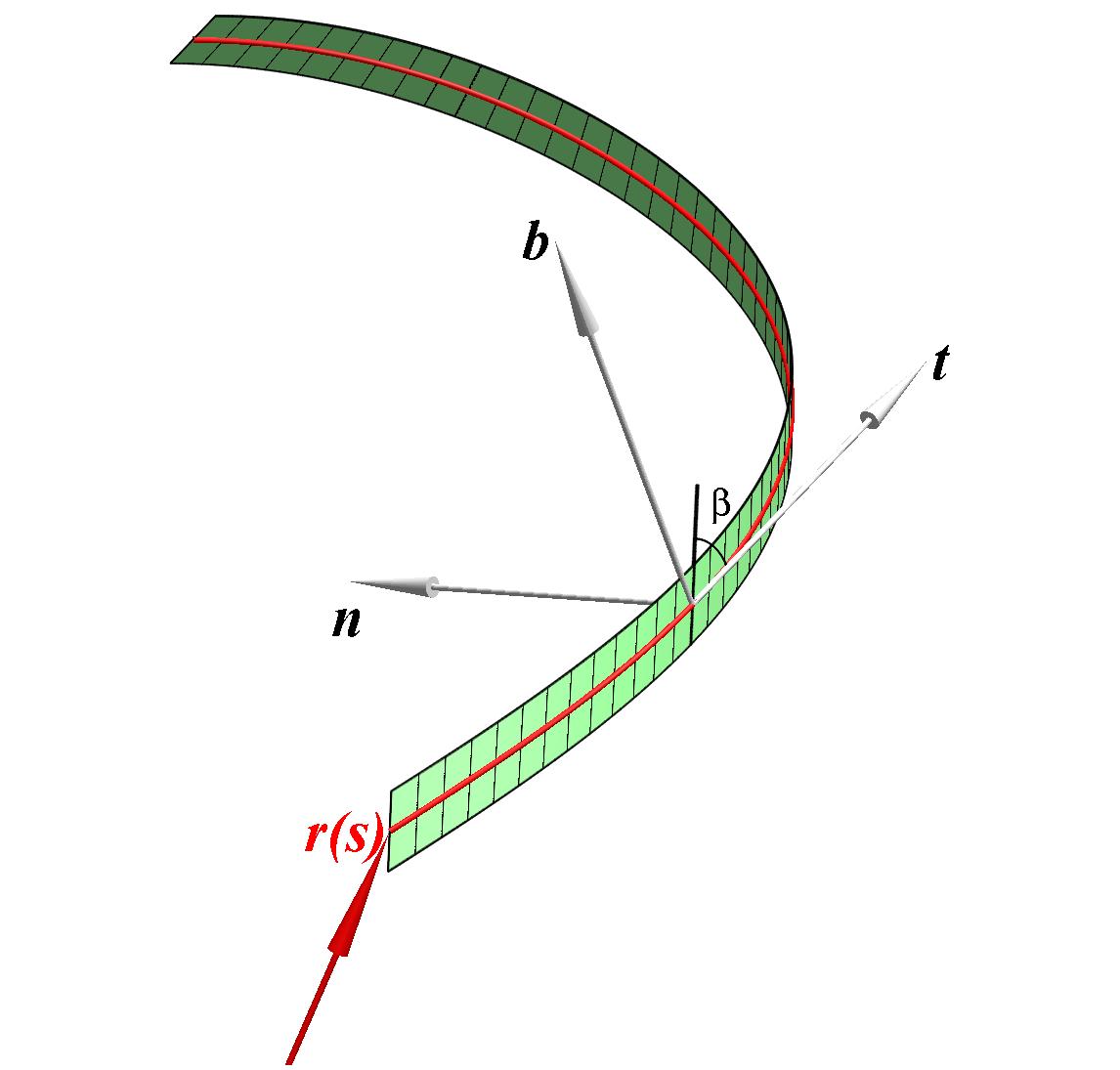}
\end{center}
\caption{A developable strip is made up of straight generators in the
rectifying plane of tangent, ${\bm t}$, and binormal, ${\bm b}$, to the
centreline, ${\bm r}$. The generators make an angle $\beta$ with the tangent.
${\bm n}$ is the principal normal.
}
\label{fig:devel}
\end{figure}

If the biopolymer is ribbonlike, i.e., much thinner than it is wide, then
the polymer essentially behaves as a thin sheet. Such sheets (e.g., paper)
tend to deform isometrically, i.e., without stretching. The deformed shape
of an intrinsically flat ribbon is therefore part of a developable surface.
Accordingly, an elastic developable strip model has been proposed for
ribbonlike filaments \cite{Giomi10,Starostin11}. Since developable surfaces
can be completely reconstructed from the strip's deformed centreline, the
problem of finding equilibrium solutions for such strips can be formulated
as a variational problem on a space curve for an energy functional in which
the width $2w$ appears merely as a parameter \cite{Wunderlich62,Starostin15}.
In the limit of a narrow strip, $w\to 0$, this functional reduces to the
Sadowsky functional \cite{Sadowsky30,Sadowsky31}
\begin{equation}
\int\kappa^2\left(1+\eta^2\right)^2 \,\mbox{d} s,
\label{eq:sadowsky}
\end{equation}
where $s$ is arclength, $\kappa$ is the curvature, $\eta=\tau/\kappa$ and
$\tau$ is the torsion of the curve. The straight generators of the surface
make an angle $\beta=\arctan(1/\eta)$ with the tangent to the centreline
(see Fig.~\ref{fig:devel}).
More precisely, the Sadowsky functional (\ref{eq:sadowsky}) is valid in
the limit $|w\eta'|\ll 1$, which means that $w$ does not have to be small
if the angle the generator makes with the centreline varies only very
gradually with arclength $s$. A strip deformed in the shape of a cylinder,
for example, which has $\eta'=0$, is described by Eq.~(\ref{eq:sadowsky})
(for arbitrary $w$).
An asymptotic analysis of the validity of functional (\ref{eq:sadowsky}),
in terms of geometrical and load parameters, is given in \cite{Chopin15}.
The Sadowsky functional originated in mechanical studies of M\"obius strips
\cite{Sadowsky30,Sadowsky31}. The functional is a singular limit of the
finite-width functional near inflection points of the centreline
\cite{Starostin15,Kirby15,Bartels15}.

There is a long line of research, stretching back to Manning's work
\cite{Manning87}, on equilibrium paths of elastic lines on curved surfaces.
Generally, a filament lying on a physical surface requires a distributed
reaction force from the surface onto the (intrinsically straight)
filament. The surface has to be stiff enough
to provide the required force, which will increase with the curvature of the
surface. These external forces acting on the filament induce
internal forces and hence stresses in the material.
For the important ideal model problem of a spherical surface, for
instance, both the WLC and RLC model require a reaction force
\cite{Manning87,Spakowitz03,Guven12a,Huynen16}.
Here we show that, remarkably, Sadowsky strips are spherical if forceless,
meaning that no distributed force is required to constrain them to a
spherical surface. So no tensile or compressive stresses need to be sustained
by the material. We like to speculate that nature may have found ways to
exploit this fact in the interaction between biofilaments and surfaces or
vesicles. By contrast, we mention the well-known fact that forceless solutions
of the Kirchhoff rod (RLC) are helices (with the straight rod and the ring as
degenerate states), while for the special case of the Euler elastica (WLC)
they are rings (or straight rods).

In fact, the Sadowsky functional is just the simplest functional of a
family of functionals whose equilibrium curves are spherical. Therefore,
in the next section we start with the more general formulation of a
geometric variational problem on a space curve.


\section{Geometric variational problems on space curves}

A space curve $\gamma$: $[0,L]\rightarrow \mathbb{R}^3$ without inflection
points is completely characterised (up to Euclidean motions) by its curvature
$\kappa(s)$ ($> 0$) and the ratio $\eta(s)=\tau(s)/\kappa(s)$, where
$\tau(s)$ is the torsion. We consider functionals on such curves of the form
\begin{equation}
U(\gamma)=\int_0^L l(\kappa,\eta)\,\mbox{d} s.
\label{eq:energy}
\end{equation}
Functionals of this type appear in a range of applications. For instance,
the classical case $l=\kappa^2$ gives the Euler elastica used as a model
for the bending of elastic rods or polymers. The case
$l=(A\kappa+B\eta)\kappa$ gives the isotropic Kirchhoff rod having both
bending and torsional stiffness \cite{Langer96}, while
$l=(A+B\eta^2)\kappa^2$ describes a thin strip whose material frame is
locked to the Frenet frame and which therefore bends only about a single
principal axis \cite{Mahadevan93}. The linear function $l=A+B\kappa+C\tau$,
meanwhile, which gives rise to generalised (Lancret) helices (having
constant $\eta$), has been proposed for protein chains \cite{Barros14}.
Functionals $U$ as in Eq.~(\ref{eq:energy}) also appear in the localised
induction hierarchy, an idealised model of the evolution of vortex
filaments in three-dimensional inviscid incompressible fluids
\cite{Langer96}, and its generalisations \cite{Perline10}. The kinematics
of space curves is furthermore related to integrable systems such as the
nonlinear Schr\"odinger equation and the modified Korteweg-de Vries
equation \cite{Lamb77,Goldstein92}.

Critical points of $U$ satisfy the following equilibrium conditions
\cite{Starostin09a}:
(a) balance equations for the components of the internal force
$\mathsf{F}=(F_t,F_n,F_b)^\intercal$ and moment
$\mathsf{M}=(M_t,M_n,M_b)^\intercal$ expressed in the Frenet frame
$\{\bm{t},\bm{n},\bm{b}\}$ (tangent, principal normal and binormal):
\begin{align}
\mathsf{F}'+\boldsymbol{\omega}\times\mathsf{F} &= \mathsf{0}, \label{eq:Fbalance} \\
\mathsf{M}'+\boldsymbol{\omega}\times\mathsf{M}+\mathsf{t}\times\mathsf{F} &= \mathsf{0},
\label{eq:Mbalance}
\end{align}
where $\boldsymbol{\omega}=(\kappa\eta,0,\kappa)^\intercal$ is the curvature
(Darboux) vector in the Frenet frame and $\mathsf{t}=(1,0,0)^\intercal$,
and (b) the `constitutive' relations
\begin{equation}
M_t = \frac{1}{\kappa}\frac{\partial {\mathit l}}{\partial \eta}, \quad\quad
M_b = \frac{\partial {\mathit l}}{\partial \kappa}-\frac{\eta}{\kappa}
\frac{\partial {\mathit l}}{\partial \eta}.
\label{eq:const_rel}
\end{equation}
The force vector is a constant vector in space and $\mathsf{F}^2$ and
$\mathsf{F}\cdot\mathsf{M}$ are first integrals of the equations
(\ref{eq:Fbalance}), (\ref{eq:Mbalance}). A further conserved quantity is
the Hamiltonian given by
\[
H=\kappa\frac{\partial {\mathit l}}{\partial \kappa}-l+F_t.
\]

The equations can alternatively be derived through Euler-Poincar\'e reduction
\cite{Starostin15} or by direct variation
\cite{Hangan05,Hornung10,Chubelaschwili10}.


\section{Forceless space curves}

We now consider the special case of forceless solutions,
$\mathsf{F}=\mathsf{0}$. For such solutions the moment vector is conserved
and the Hamiltonian becomes $H=\kappa l_{\kappa}-l$.
Generalising from some of the integrands $l$ in Eq.~(\ref{eq:energy})
reviewed above, we let $l$ be the product of two factors:
\begin{equation}
l(\kappa, \eta) = \kappa^n p(\eta),
\label{eq:l_n}
\end{equation}
where $n$ is an arbitrary number (not necessarily an integer)
and $p(\eta) \in C^3$ is an arbitrary positive function of its single
argument $\eta$.
The corresponding Hamiltonian is $H=(n-1)\kappa^n p(\eta)=h=\const$.
For $n\neq 0,1$, we have 
\begin{equation}
\kappa = \left(\frac{h}{(n-1) p(\eta)}\right)^{1/n} > 0.
\label{eq:kappa}
\end{equation}

The constitutive equations (\ref{eq:const_rel}) allow us to solve for
two components of the moment vector,
\[
M_t = \left[\frac{h}{(n-1) p}\right]^{1-1/n} p_{\eta}, \quad
M_b =  \left[\frac{h}{(n-1) p}\right]^{1-1/n} (n p - \eta p_{\eta}).
\]
The remaining component is found by differentiating $M_t$ and using the first
component of Eq.~(\ref{eq:Mbalance}):
\[
M_n = \left(\frac{h}{n-1}\right)^{1-2/n} p^{2/n-2} \left[ p p_{\eta\eta} +
\left(\frac{1}{n} -1\right) p_{\eta}^2\right]\eta'.
\]
It is easy to check that the above expressions satisfy the third component
of Eq.~(\ref{eq:Mbalance}) identically and that the second component can be
written as
\begin{align}
& A_2 \eta'' + A_1 \eta'^2 + \left(\frac{h}{n-1}\right)^{2/n} A_0 =0,
\label{eq:AAA} \\
& A_2= p^{2/n} \left[ p p_{\eta\eta}+\left(\frac{1}{n} -1\right)  p_{\eta}^2 \right], \\
& A_1= p^{2/n-1} \left[ p^2 p_{\eta\eta\eta}+\left(\frac{4}{n} -3\right) p  p_{\eta} p_{\eta\eta} + 2 \left(\frac{1}{n} -1 \right)^2  p_{\eta}^3 \right], \\
& A_0=p [(1+\eta^2) p_{\eta} - n \eta p].
\end{align}

We now recall the criterion for a curve to be spherical: \\
\noindent
{\bf Theorem}~\cite{Wong63}.
The necessary and sufficient conditions for a $C^4$ regular curve $\bm r(s)$
to lie on a sphere are
\begin{enumerate}
\item[(i)] the curvature $\kappa$ does not vanish (hence the torsion $\tau$
is defined),
\item[(ii)] there exists a $C^1$-function $f(s)$, such that
\[ 
f \tau = \left(\frac{1}{\kappa} \right)',  \quad f'+\frac{\tau}{\kappa} = 0 \ .
\]
\end{enumerate}
The curve satisfying this criterion lies on a sphere of radius
$R=\sqrt{\kappa^{-2}+f^2}$.
Note that the above theorem does not require nonvanishing torsion of the
curve.

Differentiating the expression for the curvature Eq.~(\ref{eq:kappa}) we
obtain 
\[
 \left(\frac{1}{\kappa} \right)'  =  \frac{1}{n} \left(\frac{n-1}{h}\right)^{1/n}\ p^{1/n-1} p_{\eta} \eta' .
\]
We define $f = \frac{1}{n \kappa} \left(\frac{n-1}{h}\right)^{1/n}\ p^{1/n-1} \frac{p_{\eta}}{\eta} \eta' $,
assuming that $\lim_{\eta \to 0}{\frac{p_\eta(\eta)}{\eta}}$ exists and is
finite. After substitution of $\kappa$ this becomes
$f = \frac{1}{n} \left(\frac{n-1}{h}\right)^{2/n}\ p^{2/n-1} \frac{p_{\eta}}{\eta} \eta' $.
Differentiating $f$ with respect to $s$ and inserting the result into the
equation $f'+\eta=0$, we arrive, after simplification, at a second-order
equation for $\eta$:
\begin{align}
& B_2 \eta'' + B_1 \eta'^2 + \left(\frac{h}{n-1}\right)^{2/n} B_0 =0,
\label{eq:BBB} \\
& B_2= p^{2/n-1} p_{\eta} \eta , \\
& B_1= p^{2/n-2} \left[ p (p_{\eta\eta}\eta-p_{\eta}) +\left(\frac{2}{n} -1\right)  p_{\eta}^2  \eta \right] , \\
& B_0=n \eta^3 .
\end{align}

We can now ask the question: for what $p(\eta)$ does Eq.~(\ref{eq:BBB})
coincide with Eq.~(\ref{eq:AAA})? If it does, then solutions of
Eq.~(\ref{eq:AAA}) are spherical. To answer the question, we match the
coefficients of our two equations, which gives two new equations:
\begin{align}
A_2 B_0 &= A_0 B_2,  \label{eq:A2B0} \\
A_1 B_0 &= A_0 B_1.  \label{eq:A1B0}
\end{align}
These are two nonautonomous ordinary differential equations for $p(\eta)$.
 
Eq.~(\ref{eq:A2B0}) simplifies to 
 \[
 \eta^2 p p_{\eta\eta}  - \left(\eta^2 + \frac{1}{n} \right) p_{\eta}^2 + \eta p p_{\eta} = 0 .
 \]
Its general solution is 
 \[
 p(\eta) = C \left(\eta^2 + \frac{N}{n} \right)^N,
 \]
where $C$ and $N$ are integration constants.
Note that the ratio $\frac{p_\eta(\eta)}{\eta}$ is well defined for $\eta=0$.
Direct substitution of the above $p(\eta)$ into the second condition
Eq.~(\ref{eq:A1B0}) reveals that it is satisfied only for $N=n$ (the
arbitrary prefactor constant $C$ is clearly of no importance). Thus, we
conclude that all forceless inflection-free minimisers of the functional
$l(\kappa, \eta) = \kappa^n p(\eta)$, $n\neq 0,1$, are spherical only for
\begin{equation}
l(\kappa, \eta) = C \kappa^n (1+\eta^2)^n, \quad\quad C=\const.
\label{eq:l_spherical}
\end{equation}
The radius of the sphere is
$R = \left| \frac{n-1}{n} \frac{M}{h} \right|$, where $M^2=M_t^2+M_n^2+M_b^2>0$.
A special analysis reveals that for $n=1$, Eq.~(\ref{eq:l_spherical}) gives,
among other solutions, arbitrary planar curves ($\eta=0$). For $n=0$,
Eq.~(\ref{eq:l_spherical}) is trivial, but Eq.~(\ref{eq:l_n}) gives
Lancret helices, for arbitrary nonconstant $p$.


\section{The Sadowsky functional -- forceless strip solutions}

For $n=2$ in Eq.~(\ref{eq:l_spherical}) we obtain the Sadowsky functional
Eq.~(\ref{eq:sadowsky}):
\begin{equation}
U_S(\gamma)=\int\nolimits_{0}^{L}\kappa^2\left(1+\eta^2\right)^2 \,\mbox{d} s.
\label{eq:energyNarrow}
\end{equation}
For forceless strips Eqs~(\ref{eq:Fbalance}), (\ref{eq:Mbalance}) and
(\ref{eq:const_rel}) reduce to
\begin{align}
& M_t'=\kappa M_n, \quad\quad M_n'=\kappa\eta M_b-\kappa M_t, \quad\quad
M_b'=-\kappa\eta M_n, \label{eq1} \\
& M_t=4\kappa\eta(1+\eta^2), \quad\quad M_b=2\kappa(1-\eta^4), \label{eq2}
\end{align}
while the Hamiltonian is
\begin{equation}
H=\kappa^2(1+\eta^2)^2.
\label{ham}
\end{equation}

The remaining normal component of the moment may be found from the first
(or third) equation in (\ref{eq1}) and (\ref{ham}):
\begin{equation}
M_n = 4 (1+\eta^2) \eta'.
\end{equation}
Combination with the second equation in (\ref{eq1}) and again (\ref{ham})
then gives
\begin{equation}
2(1+\eta^2) \eta'' + 4 \eta \eta'^2 + h \eta =0,
\label{eq:nrw_rect_eta}
\end{equation}
where $h$ is the value of the Hamiltonian.
The theorem above tells us that solutions of this equation represent
spherical curves, i.e., centrelines of narrow forceless rectangular
strips are spherical curves. The radius of the sphere equals
$R=\frac{M}{2h}$.

Integrating Eq.~(\ref{eq:nrw_rect_eta}) once gives the moment first integral
\begin{equation}
G(\eta,\eta') := 4 (1+\eta^2)^2 (4 \eta'^2  + h) = M^2.
\label{eq:nrw_rect_eta1int}
\end{equation}
Analysis of the derivatives of $G(\eta,\eta')$ reveals that there always
exists only one critical point at the origin and that it is always a centre
point. Therefore, all the orbits in the phase plane are closed
(see Fig.~\ref{fig:fig_phase}).

\begin{figure}
\begin{center}
\includegraphics[width=0.5\textwidth]{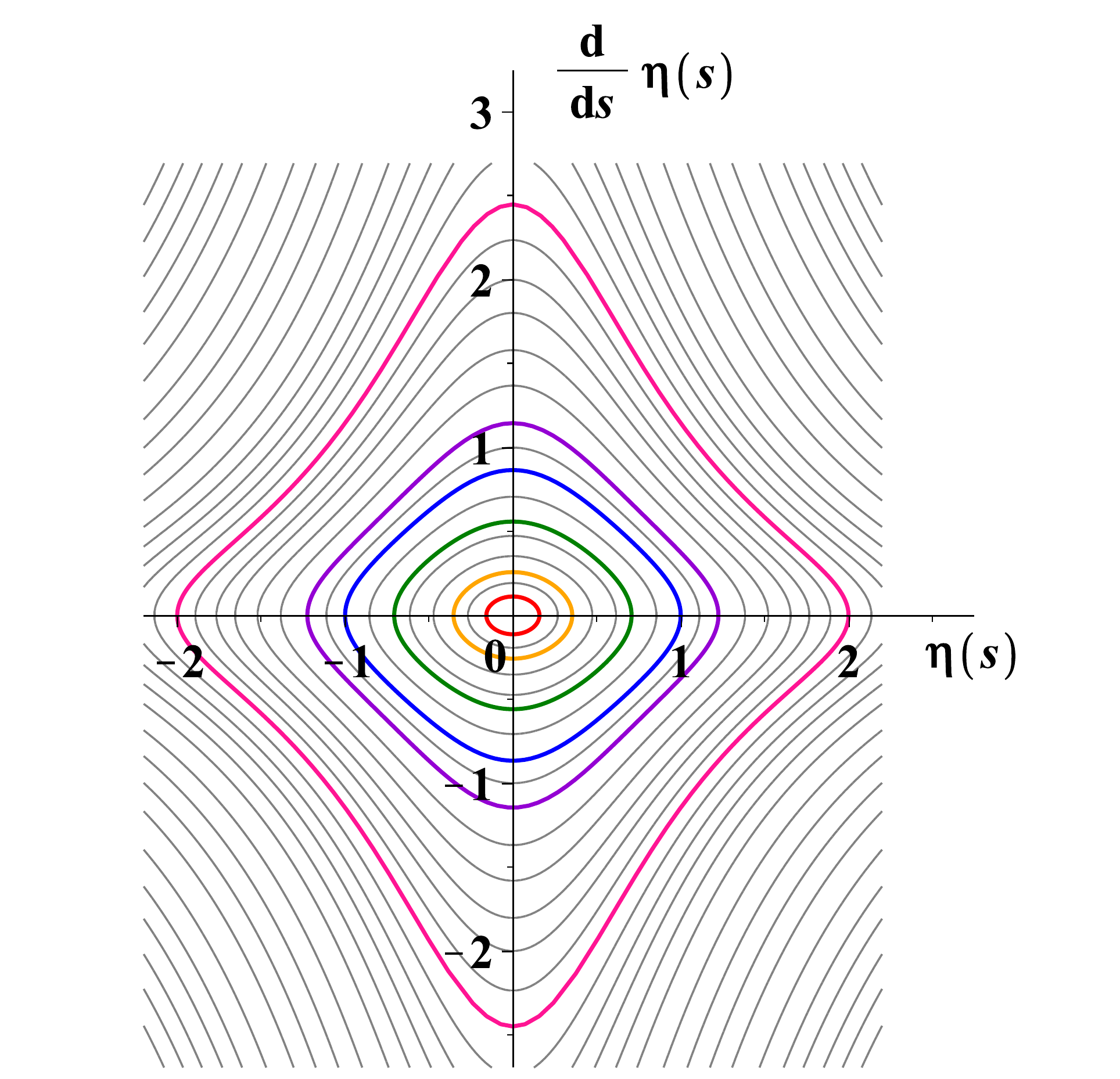}
\end{center}
\caption{Phase portrait for Eq.~(\ref{eq:nrw_rect_eta1int}) with orbits for
$M=2.05$, 2.25, 3, 4, 5 and 10 (inner to outer) highlighted ($h=1$).
}
\label{fig:fig_phase}
\end{figure}

Further integration of Eq.~(\ref{eq:nrw_rect_eta1int}) yields
\[
\pm \frac{2}{\sqrt{h}} \int_0^\eta \frac{1+\eta^2}{\sqrt{A^2 -
(1+\eta^2)^2 }} \mbox{d} \eta = s-s_0 \ ,
\]
where $A^2 = \frac{M^2}{4h} = hR^2 \ge 1$, the inequality following from
Eq.~(\ref{eq:nrw_rect_eta1int}).
Evaluation of the integral delivers the final equation
\begin{align}
\sqrt{2 A} \left[ 2 \E \left(\eta \sqrt{\frac{2A}{(A-1)(A+1+\eta^2)}}, \sqrt{\frac{A-1}{2A}} \right) 
- \F \left(\eta \sqrt{\frac{2A}{(A-1)(A+1+\eta^2)}}, \sqrt{\frac{A-1}{2A}} \right) \right] - \nonumber \\
- 2 \eta \sqrt{\frac{A-1-\eta^2}{A+1+\eta^2}} = \pm \sqrt{h} (s-s_0) \ ,
\label{solution}
\end{align}
where $\F(z,k)  = \int_0^z (1-k^2 \sin^2 u)^{-\frac12} \,\mbox{d} u$ and 
$\E(z, k) = \int_0^z (1-k^2 \sin^2 u)^{\frac12} \,\mbox{d} u$ are the
incomplete elliptic integrals of the first and second kind, respectively
(with $k$ the elliptic modulus), and $s_0$ is an integration constant. Once
this equation is solved for $\eta$, the curvature can be computed as
\[
\kappa=\frac{\sqrt{h}}{1+\eta^2}.
\]

As follows from Eq.~(\ref{eq:nrw_rect_eta1int}), $\eta'$ goes through a
maximum or minimum when $\eta=0$, while $\eta$ goes through a maximum or
minimum, $\eta=\pm\sqrt{A-1}$, when $\eta'=0$. Using this, the period can be
computed from Eq.~(\ref{solution}) as
\[
T=4\sqrt{\frac{2 A}{h}} \left[ 2 \E \left(\sqrt{\frac{A-1}{2A}} \right) 
- \K \left(\sqrt{\frac{A-1}{2A}} \right) \right] ,
\]
where $\K(k)$ and $\E(k)$ are the complete elliptic integrals of the first
and second kind, respectively.
The curvature is then periodic with period $T/2$.
The expression for the Hamiltonian implies that zeroes of $\eta$ correspond
to maxima of the curvature, $\kappa_{max} = \sqrt{h}$, while
$\eta$ has extrema at points where $\kappa$ has a minimum,
$\kappa_{min} = \frac{\sqrt{h}}{A} = \frac{1}{R}$
(see Figs.~\ref{fig:fig_m1} and \ref{fig:fig_m2}). Note that the torsion
$\tau$ averaged over a period $T$ is zero. Solutions are therefore achiral.

We also note that the tangential component of the moment is proportional
to $\eta$: $M_t=4 \sqrt{h} \eta$. Thus the tangent to the centreline makes
an angle with the moment vector with cosine equal to $M_t/M=2\eta/A$.
This implies that the tangent to the centreline is oriented orthogonally to
the fixed axis of the moment vector at points where $\eta=0$,
while the tangent to the centreline is aligned with the moment vector at
points where $\eta=\pm A/2$.
Since $-\sqrt{A-1} \le \eta \le \sqrt{A-1}$, the latter occurs at maximum
$|\eta|$ if $A=2$, i.e., $\eta=\pm 1$ (Fig.~\ref{fig:fig_m2}a gives
an example for $h=1$, $M=4$).

Shapes of strips on the sphere are shown in Figs.~\ref{fig:fig_m1} and
\ref{fig:fig_m2}. Here the strips are drawn with a small width to illustrate
that they rotate relative to the (imaginary) spherical surface. The
angle $\chi$ between the normal to the developable surface of the ribbon
at its centreline and the normal to the sphere can be found from the equation
$\kappa \cos\chi = \frac{1}{R}$. We see that at points of vanishing $\eta$,
where the generator is orthogonal to the centreline, this angle
reaches its maximum value, while it vanishes at points of maximum
$|\eta|$. In the latter case the tangent plane to the ribbon's surface is
also tangent to the sphere.

Strips are generally not closed on the sphere, but periodic boundary
conditions (in both space and curvature)
could be imposed, which would fix one of the two free
parameters $(M,h)$, leaving a one-parameter family of closed solutions.
Note that these structures would be closed as a strip since periodicity of
curvature and torsion enforces periodicity of the Frenet frame and alignment
of the end generators. They would have high-order spatial symmetry, namely
$D_{nd}$ symmetry ($n$ being a mode number), with planes of reflection symmetry
through the moment vector alternating with axes of $\pi$-rotation symmetry
perpendicularly intersecting the central moment axis and transversely
intersecting the symmetry planes. Non-closed (quasi-periodic) strip solutions,
meanwhile, have $D_{\infty h}$ symmetry.
Structures with either of these symmetry groups must indeed have zero force
as there can neither be a force component in a plane of reflection symmetry
nor along an axis of rotation symmetry.

\begin{figure}[p]
\begin{center}
{\bf~(a)~~~~~~~~~~~~~~~~~~~~~~~~~~~~~~~~~~(b)~~~~~~~~~~~~~~~~~~~~~~~~~~~~~~~~~~(c)~}
\vspace{0.2cm} \\
\includegraphics[height=0.22\textwidth]{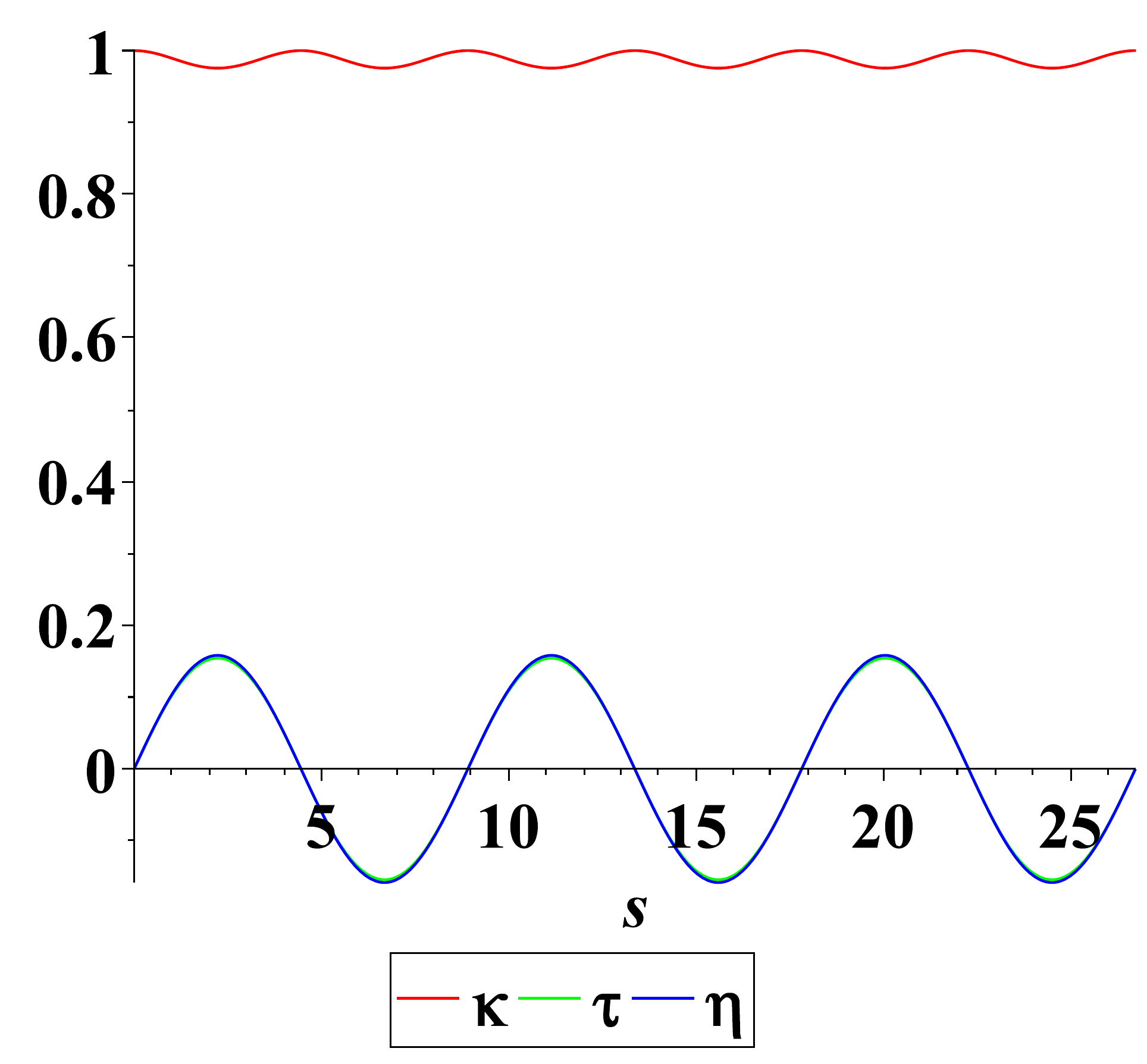}
\hspace{1.5cm}
\includegraphics[height=0.22\textwidth]{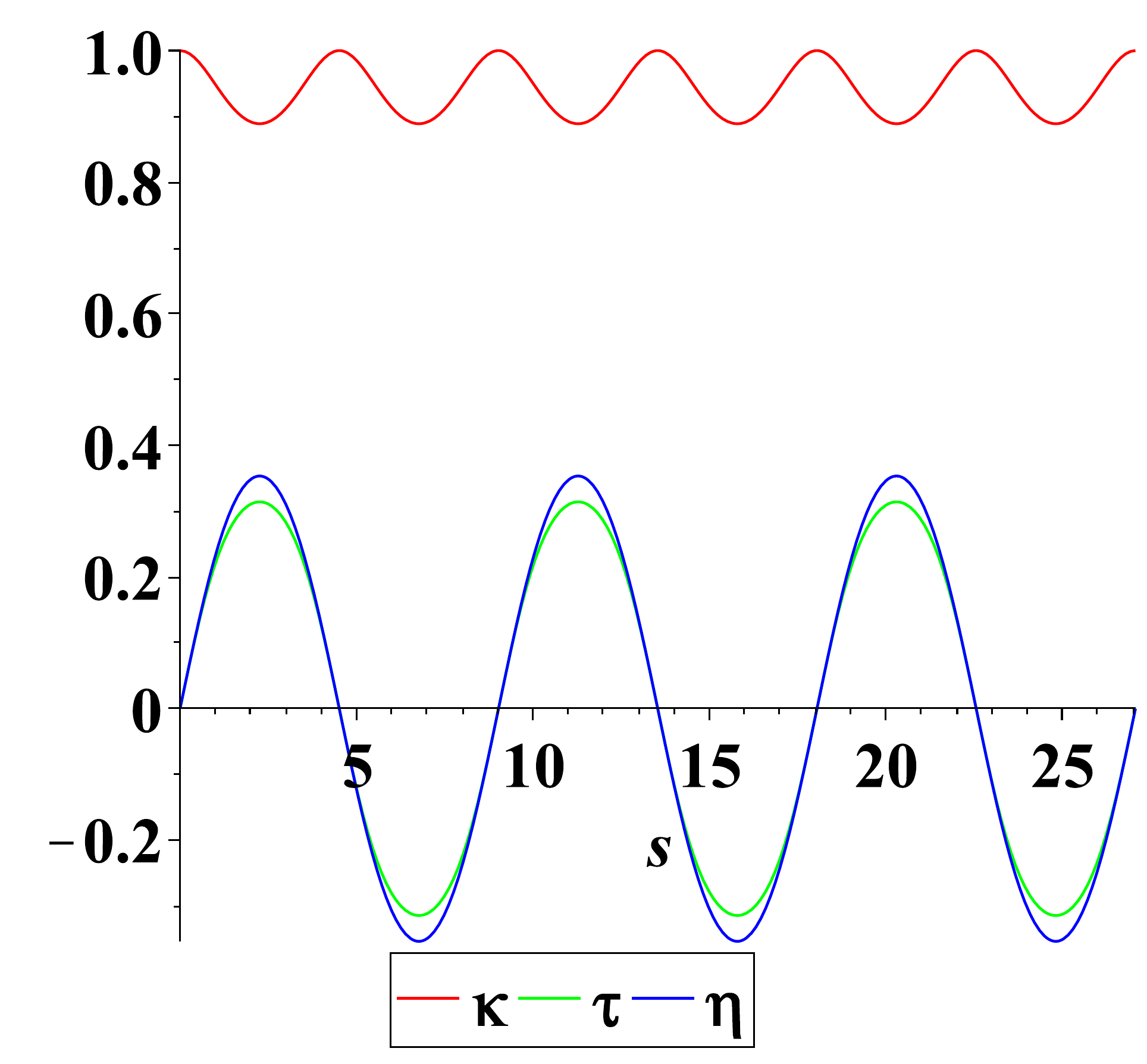}
\hspace{1.5cm}
\includegraphics[height=0.22\textwidth]{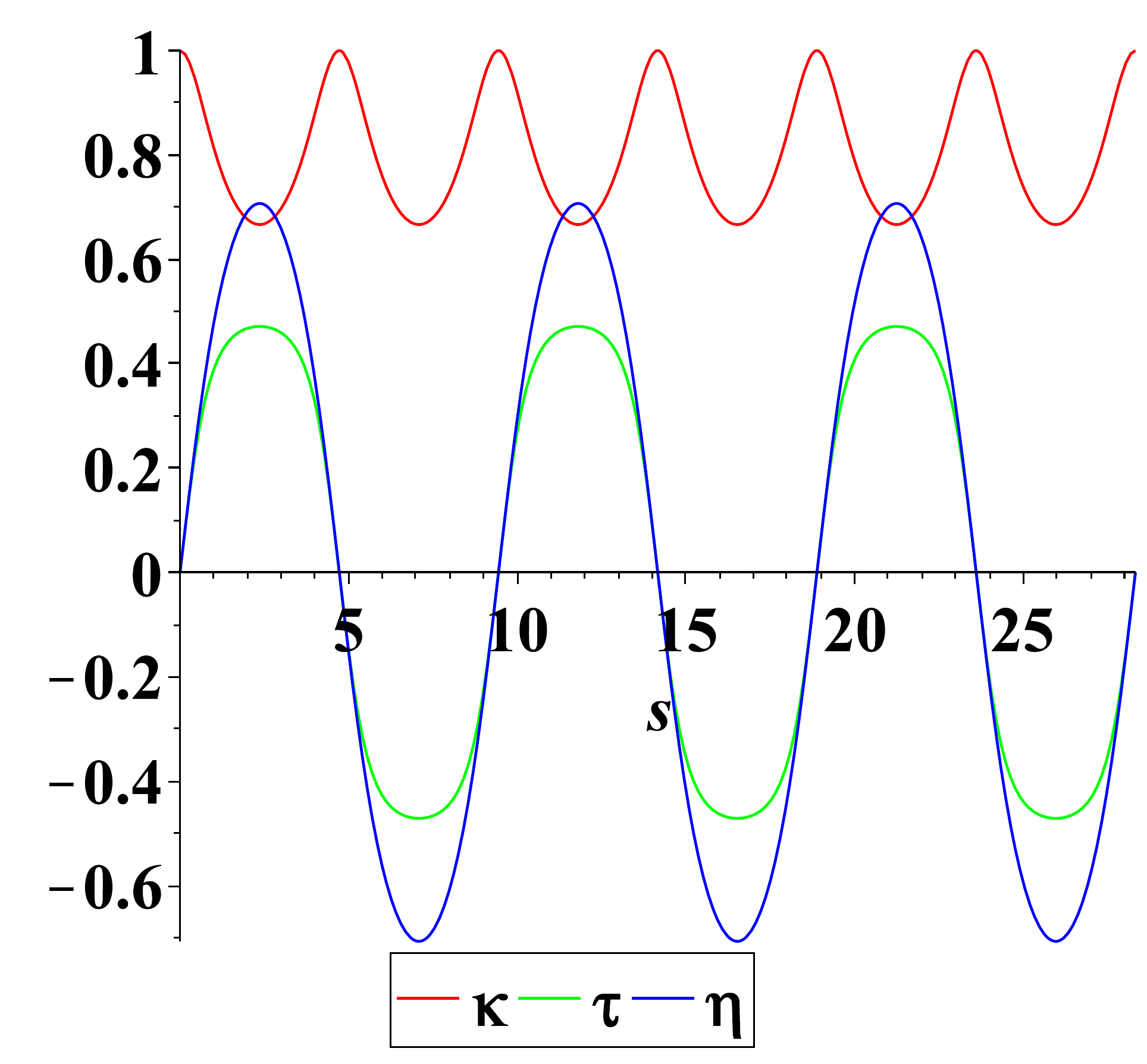} \\
\vspace{0.2cm}
\includegraphics[height=0.22\textwidth]{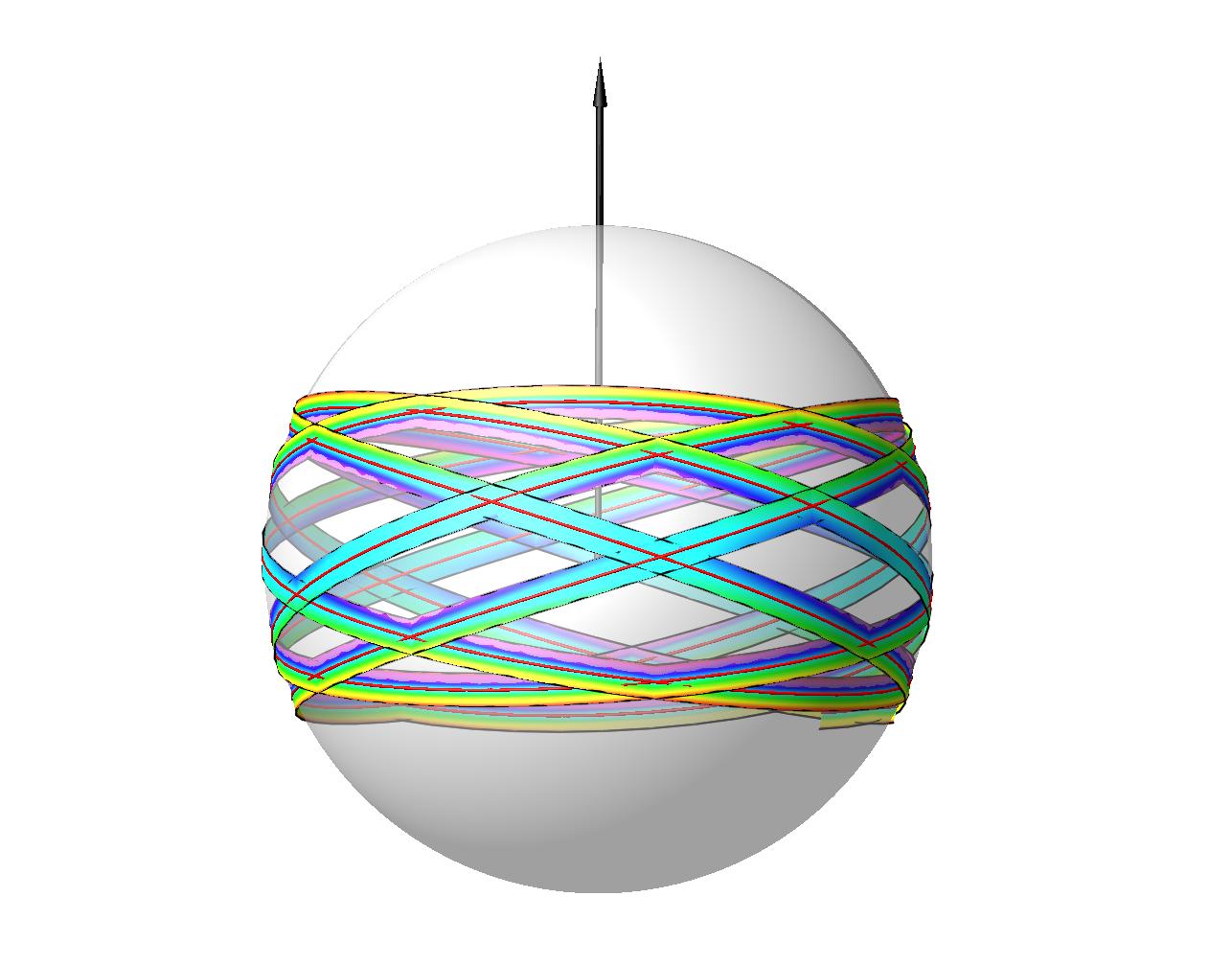}
\hspace{1cm}
\includegraphics[height=0.22\textwidth]{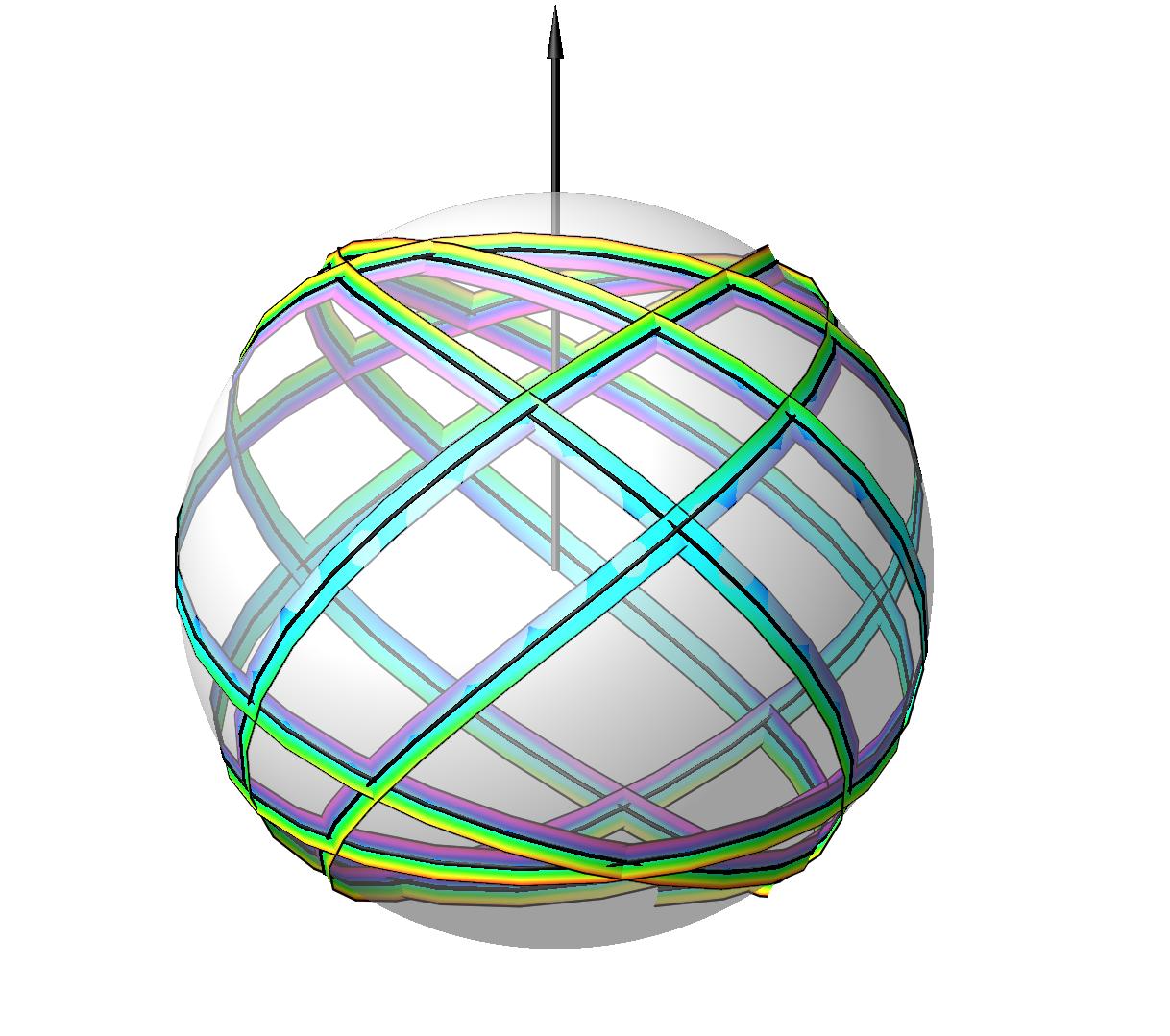}
\hspace{1cm}
\includegraphics[height=0.22\textwidth]{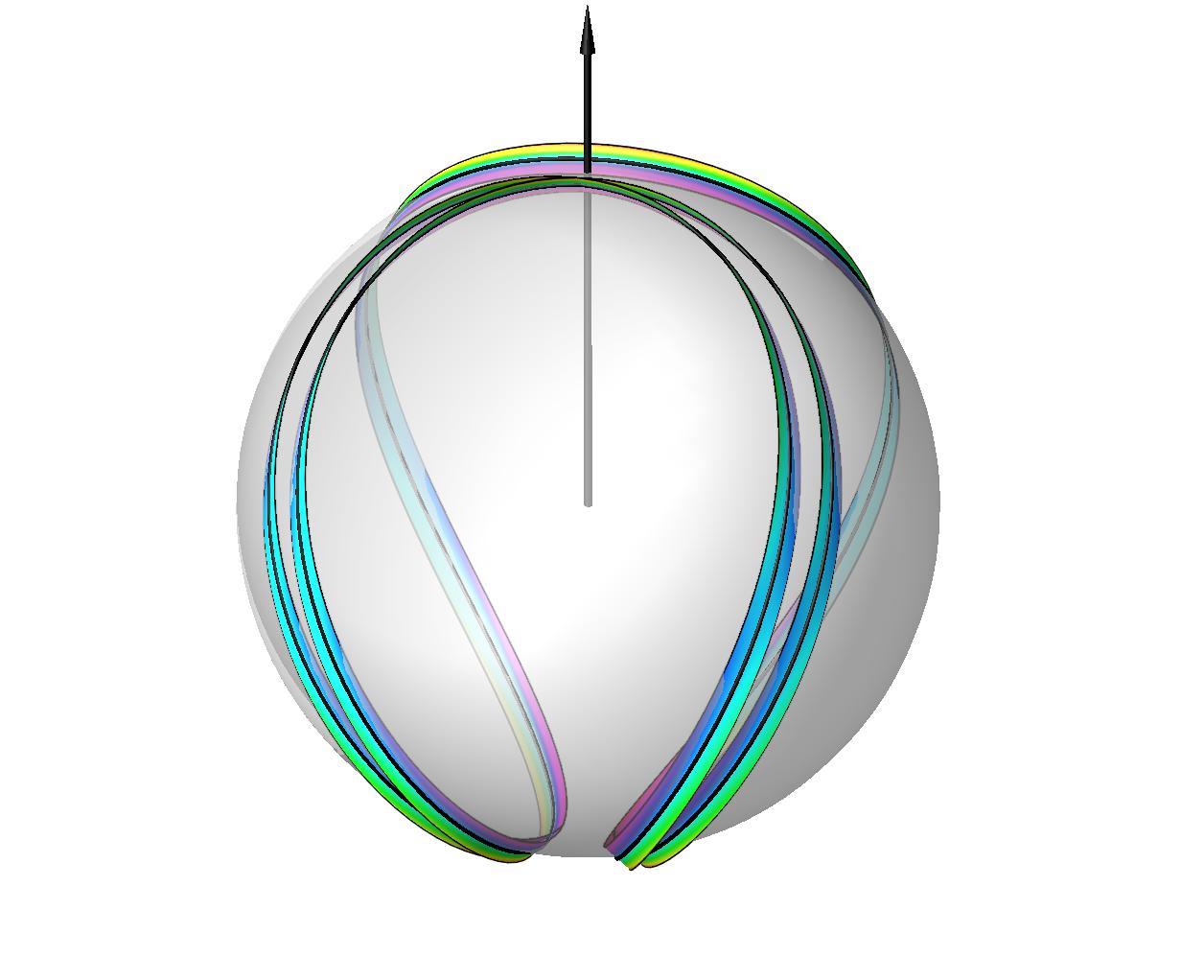}
\end{center}
\vspace{-0.6cm}
\caption{Forceless Sadowsky strip solutions. (Top) Curvature $\kappa(s)$,
torsion $\tau(s)$ and their ratio $\eta(s)$, $s \in [0, 3T]$, for
(a) $M=2.05$ ($T = 8.91355$),  
(b) $M=2.25$ ($T = 9.02503$), 
(c) $M=3$ ($T = 9.44378$).  
(Bottom) Corresponding spherical shapes for $s \in [0, 5T]$.
The black arrow indicates the moment vector. ($h=1$.)
}
\label{fig:fig_m1}
%
\begin{center}
{\bf~(a)~~~~~~~~~~~~~~~~~~~~~~~~~~~~~~~~~~(b)~~~~~~~~~~~~~~~~~~~~~~~~~~~~~~~~~~(c)~}
\vspace{0.2cm} \\
\includegraphics[height=0.22\textwidth]{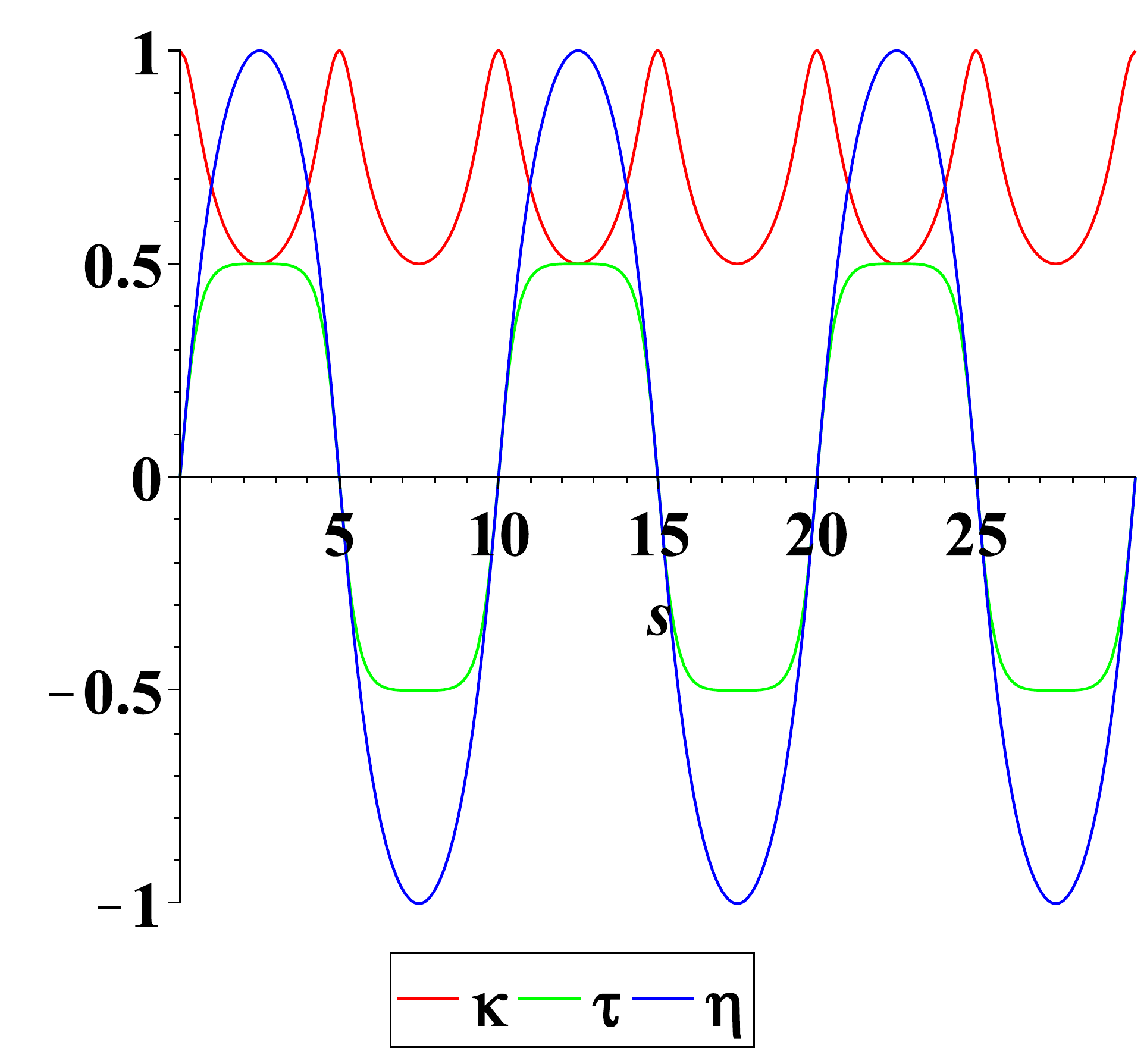}
\hspace{1.5cm}
\includegraphics[height=0.22\textwidth]{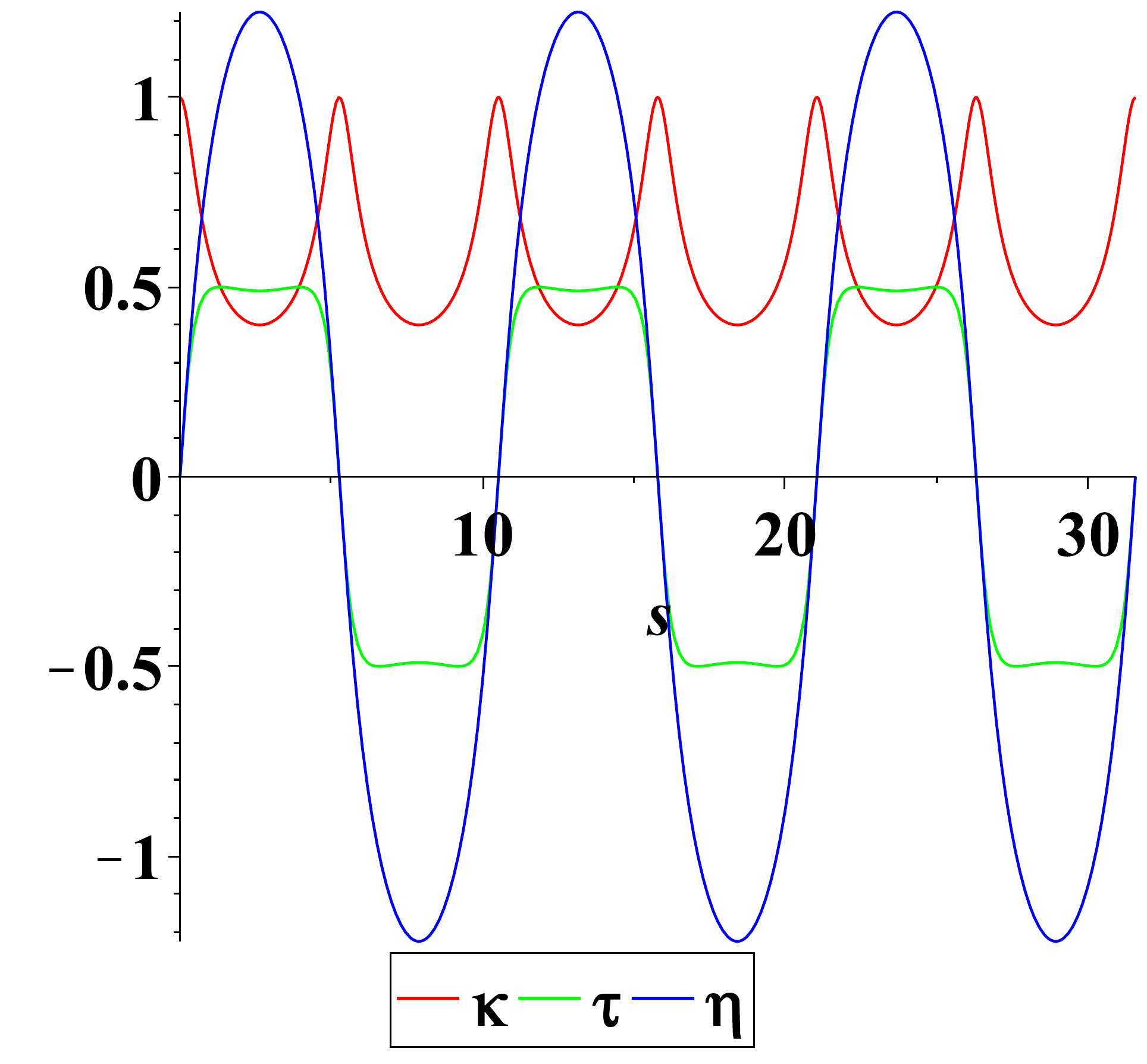}
\hspace{1.5cm}
\includegraphics[height=0.22\textwidth]{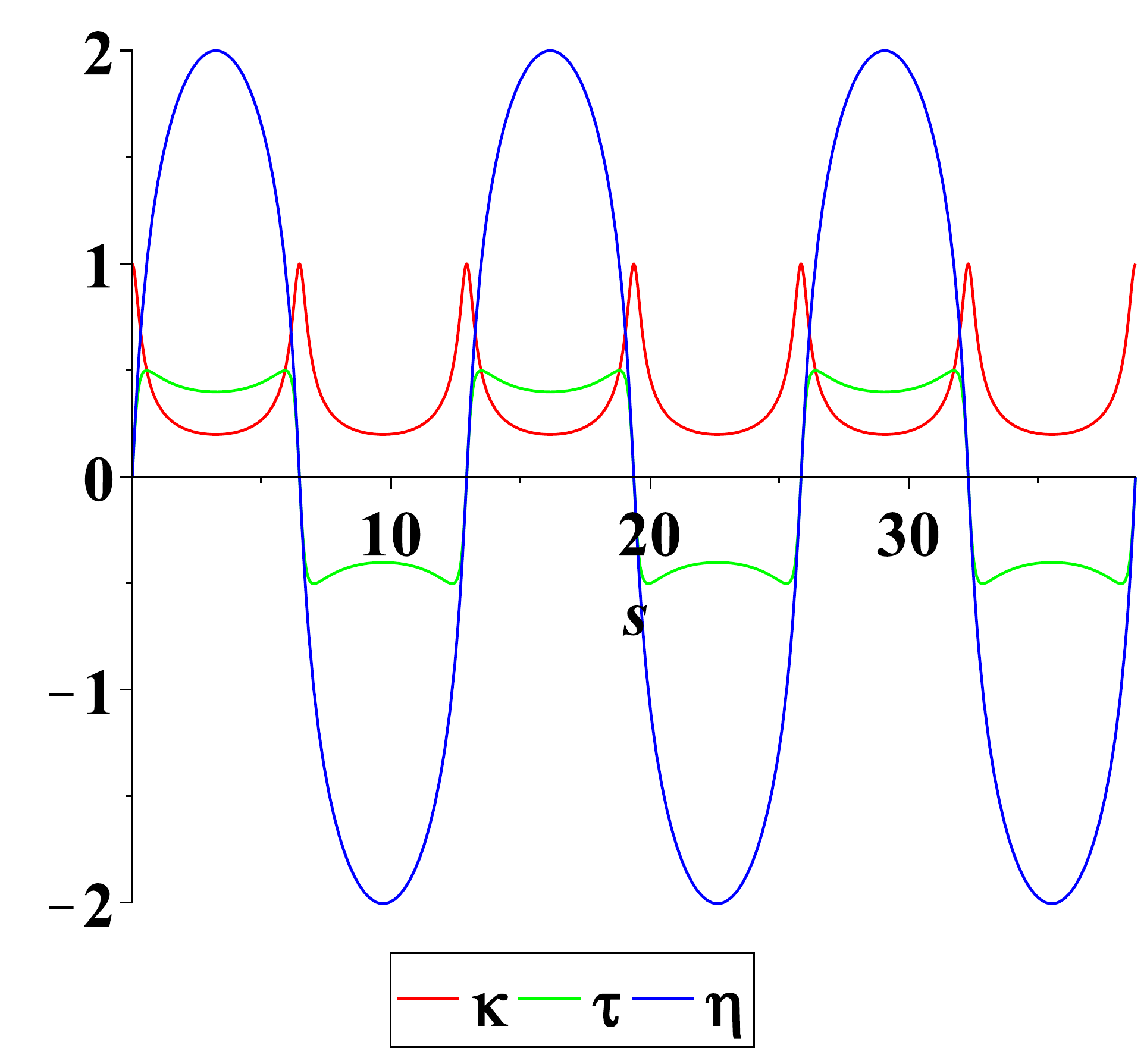} \\
\vspace{0.2cm}
\includegraphics[height=0.22\textwidth]{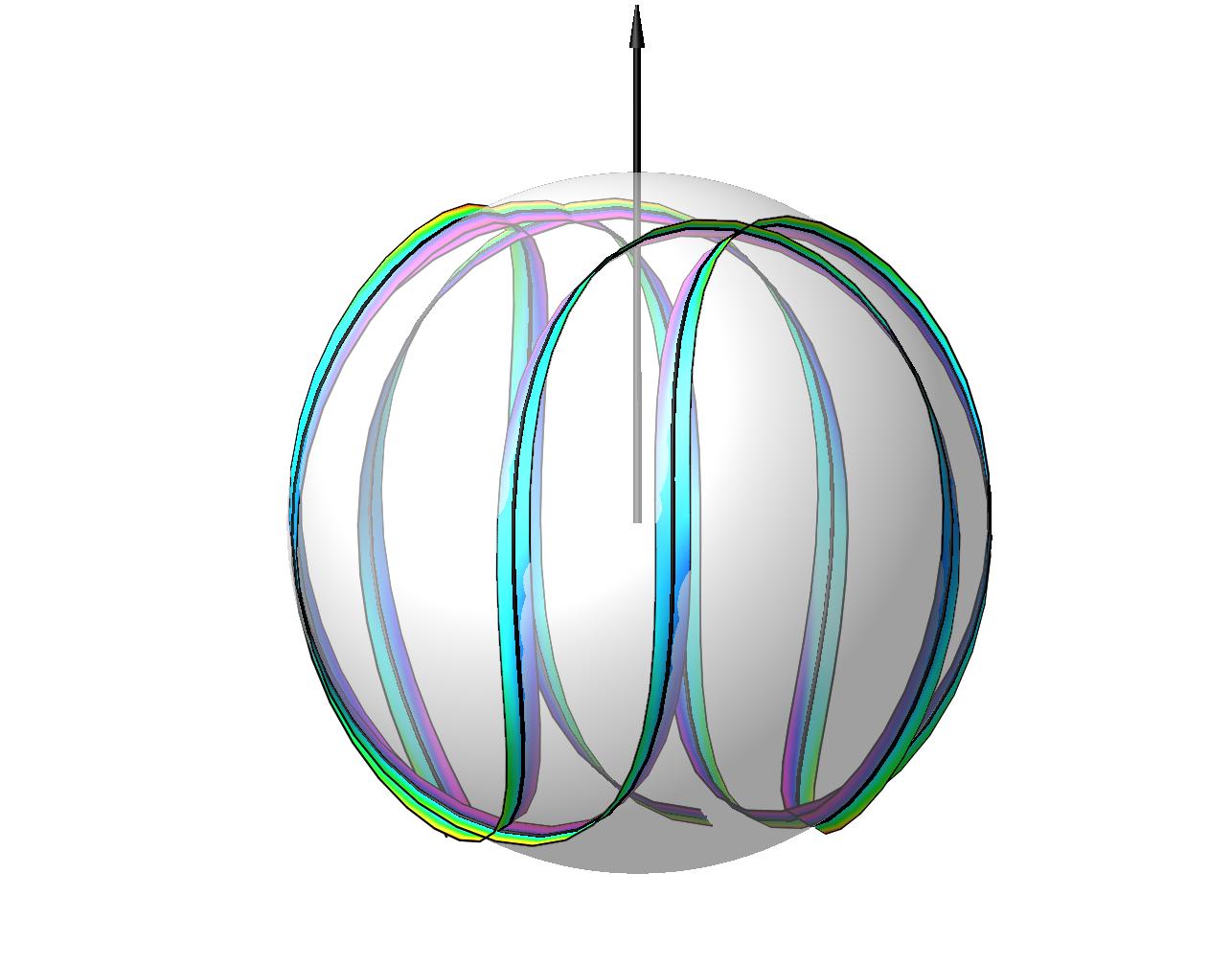}
\hspace{1cm}
\includegraphics[height=0.22\textwidth]{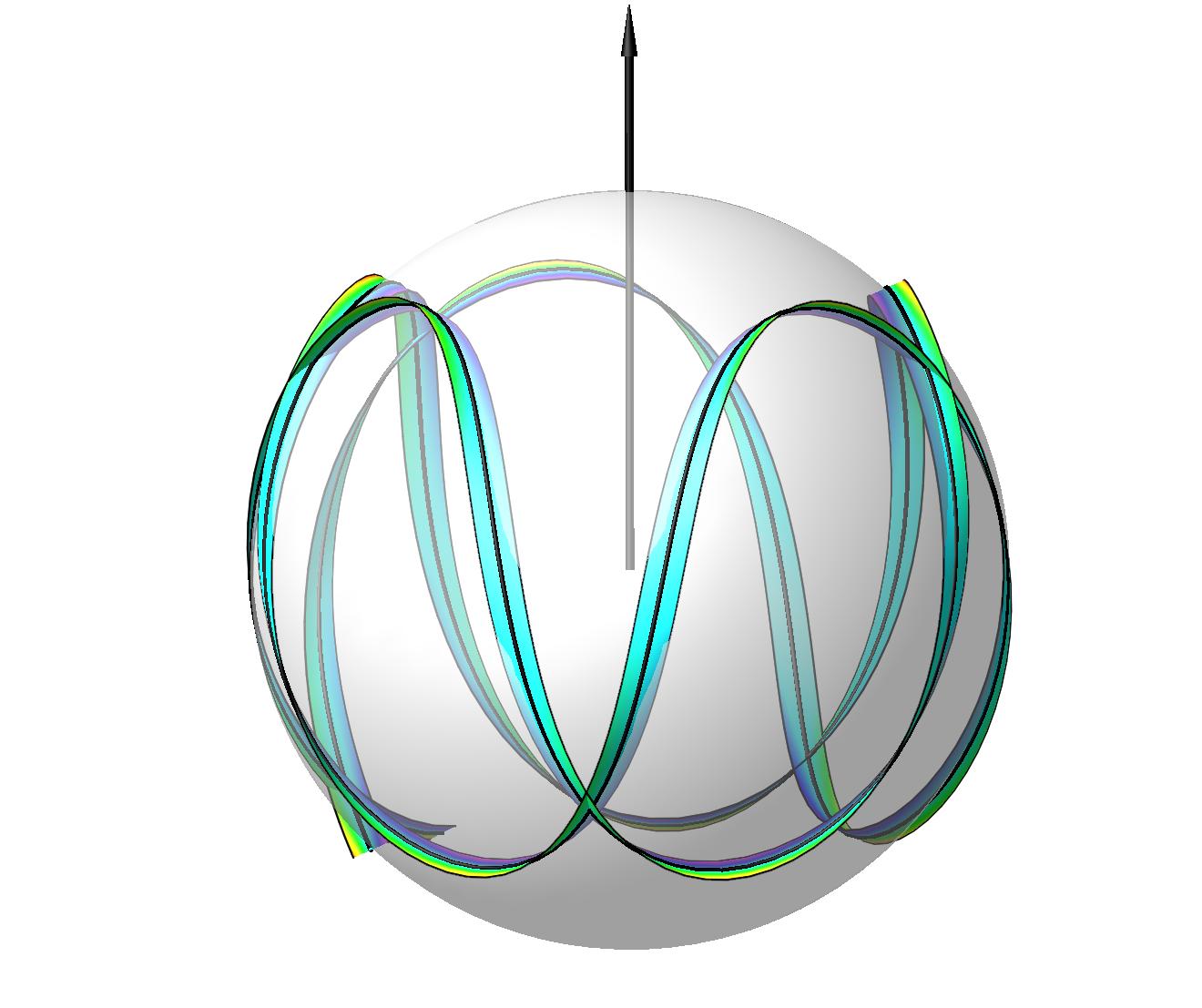}
\hspace{1cm}
\includegraphics[height=0.22\textwidth]{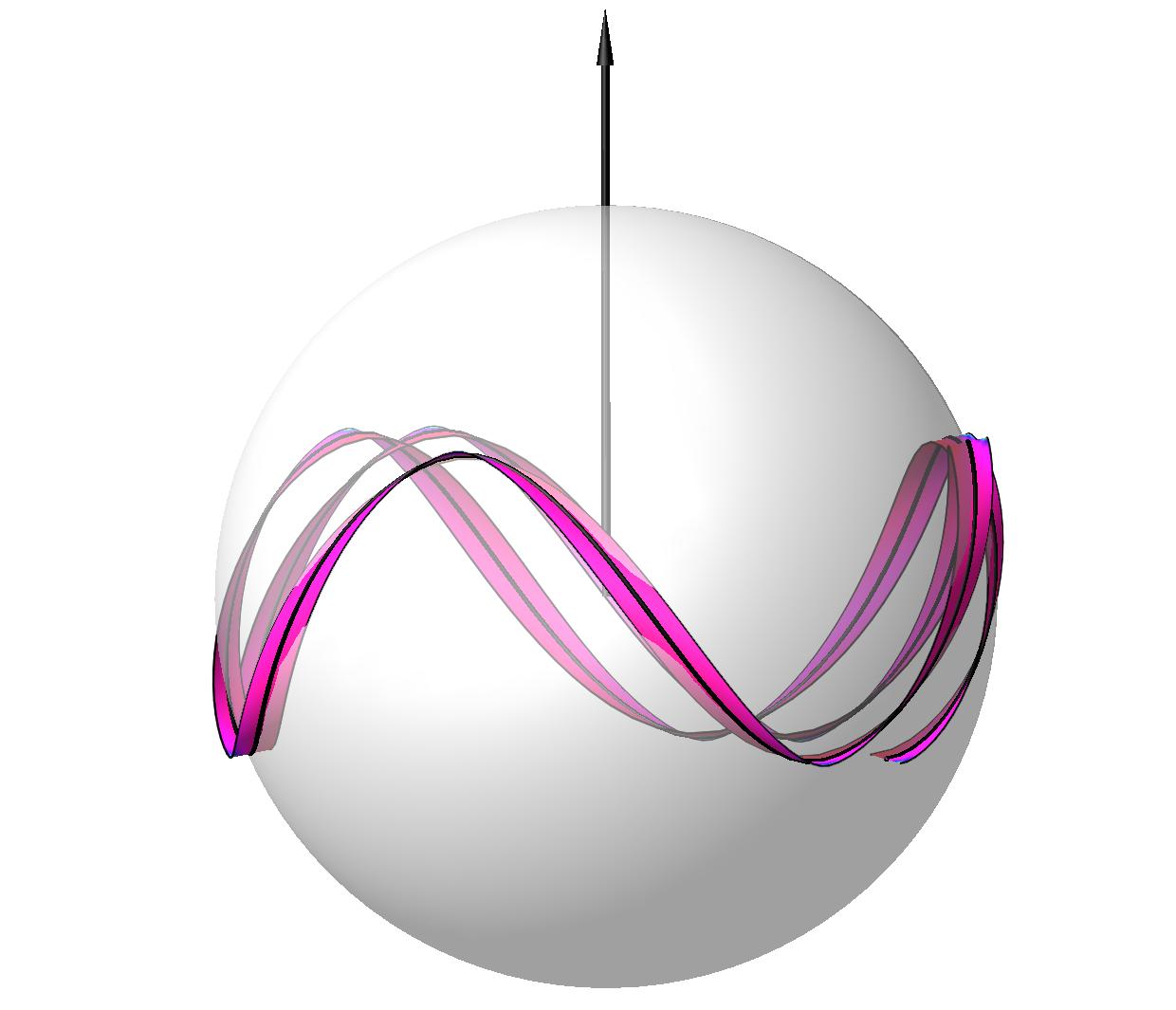}
\end{center}
\vspace{-0.6cm}
\caption{Continued from Fig.~\ref{fig:fig_m1} for
(a) $M=4$ ($T = 9.99339$),  
(b) $M=5$ ($T = 10.52595$), 
(c) $M=10$ ($T = 12.91809$).  
}
\label{fig:fig_m2}
\end{figure}


\section{Discussion}

We have shown that a class of energy functionals for elastic filaments,
which includes the Sadowsky energy for a narrow strip, has spherical
forceless extremals. For the Sadowsky case solutions depend on two
parameters, the values of the two first integrals, i.e., the magnitude of
the moment ($M$) and the Hamiltonian ($h$), which is also the (normalised)
bending energy density. The radius of the sphere is $\frac{M}{2h}$.

The class of functionals with this property may be wider.
However, it does not include the corrected Sadowsky functional constructed
in \cite{Freddi16} (although this correction only affects solutions where
$|\eta|>1$, so solutions for which $|\eta|\leq 1$ everywhere are still
spherical). Nor does it include the narrow limit ($w\to 0$) of the functional
for annular strips derived in \cite{Dias15}, nor, seemingly, the functional
for narrow residually-stressed strips derived in \cite{Efrati15}.
It would be interesting to find all functionals of the form (\ref{eq:energy})
(or, more generally, functionals with $l=l(\kappa,\eta,\kappa',\eta',...)$
\cite{Starostin09a}) with unconstrained spherical solutions, analogous to
all functionals with forceless helical solutions having been characterised
in \cite{Barros14}.

We stress that in this paper we have not considered any constraint on the
strip. In particular, the surface of the strip is not required to lie in the
surface of the sphere, although solutions, as in Fig.~\ref{fig:fig_m1}a,
that remain close to the equator (i.e., have small geodesic curvature),
rotate out of the surface only very little.
Strips adhered to a spherical surface (similar to the growing crystals
studied in \cite{Meng14}) would obviously have Gaussian curvature $1/R^2$,
with $R$ the radius of the sphere. The surface of the strip would then not
be developable and therefore not be described by the Sadowsky functional.
However, the Sadowsky functional can still be expected to provide a good
approximation for the mechanics of a physical ribbon if the stretching
energy $U_s$ is much smaller than the bending energy $U_b$.
Now, for an adhered ribbon whose geodesic curvature is much smaller than its
normal curvature, we estimate $U_s\sim t(w/R)^4$ and $U_b\sim t^3/R^2$, where
$t$ is the thickness of the ribbon (both energies per unit area). We thus
require $w/R\ll t/w$ (in addition to $t/w\ll 1$ for any ribbon model) and we
conclude that the (approximate) validity of the Sadowsky model for such
adhered spherical ribbons does not extend to arbitrarily thin ribbons.


\begin{acknowledgments}
We thank an anonymous referee for insightful comments that helped us to
improve the paper.
\end{acknowledgments}

\addcontentsline{toc}{section}{References}

\bibliographystyle{plain}

\end{document}